\newcount\draft\draft=0 % set to 0 for submission
\newcount\cameraready\cameraready=1 % set to 1 for publication

\documentclass[letterpaper,twocolumn,10pt]{article}
\usepackage{usenix}

\usepackage{amsmath}
\usepackage{bm} % bold math
\usepackage[font=small,labelfont=bf]{caption}
\usepackage{graphicx}
\usepackage{lastpage}
\usepackage{listings}
\usepackage{mathptmx}
\usepackage{paralist}
\usepackage{url}
\usepackage{xspace}
\usepackage{xxx}

\definecolor{mygreen}{rgb}{0,0.6,0}
\definecolor{mygray}{rgb}{0.5,0.5,0.5}
\definecolor{mymauve}{rgb}{0.58,0,0.82}

\DeclareGraphicsExtensions{.pdf,.jpg,.png}
\graphicspath{{./R/}{./figs/}}

\newcommand{\sysname}{SAP\xspace}
\newcommand{\sysnamefull}{Selective Approximate Protocol\xspace}

\newcommand{\secref}[1]{Section~\ref{#1}}

\newcommand{\figref}[1]{Figure~\ref{#1}}
\newcommand{\tblref}[1]{Table~\ref{#1}}
\newcommand{\etal}{\mbox{et al.}\xspace}
\newcommand{\mypara}[1]{\textbf{#1}}
\newcommand{\term}[1]{\textit{#1}}
\newcommand{\lil}{\lstinline}

\begin{document}

%don't want date printed
\date{}

%make title bold and 14 pt font (Latex default is non-bold, 16 pt)
\title{
\sysname: an Architecture for Selectively Approximate Wireless Communication
}

\ifnum\cameraready=1
%for single author (just remove % characters)
\author{
{\rm Benjamin Ransford}\\
Virta Laboratories, Inc.
\and
{\rm Luis Ceze}\\
University of Washington
% copy the following lines to add more authors
% \and
% {\rm Name}\\
%Name Institution
} % end author
\else
\author{}
\fi

\frenchspacing

\maketitle

% Use the following at camera-ready time to suppress page numbers.
% Comment it out when you first submit the paper for review.
\thispagestyle{empty}

\subsection*{Abstract}

Integrity checking is ubiquitous in data networks, but not all network traffic
needs integrity protection.  Many applications can tolerate slightly damaged
data while still working acceptably, trading accuracy versus efficiency to save
time and energy.  Such applications should be able to receive damaged data if
they so desire.  In today's network stacks, lower-layer integrity checks discard
damaged data regardless of the application's wishes, violating the End-to-End
Principle.  This paper argues for \emph{optional} integrity checking and gently
redesigns a commodity network architecture to support integrity-unprotected
data.  Our scheme, called \sysnamefull (\sysname), allows applications to
coordinate multiple network layers to accept potentially damaged data. Unlike
previous schemes that targeted video or media streaming, \sysname is generic.
\sysname's improved throughput and decreased retransmission rate is a good match
for applications in the domain of \emph{approximate computing}.

Implemented atop WiFi as a case study, \sysname works with existing physical
layers and requires no hardware changes.  \sysname's benefits increase as
channel conditions degrade.  In tests of an error-tolerant file-transfer
application over WiFi, \sysname sped up transmission by about 30\% on average.

\section{Introduction}\label{sec:intro}

% context: increasing wireless comms, poor channels leading to poor performance
% when we try to get everything right
Today's predominant network protocols guarantee that data will be received
exactly as it was sent, intact and usually in order.  Even UDP, which allows
datagram loss and reordering, typically runs on integrity-protected MAC and
physical layers such as Wi-Fi.

This guarantee is necessary for applications that wish to abstract away
communication details.
However, as the population of endpoints shifts from
reliable wired networks to error-prone wireless networks such as Wi-Fi and
cellular, applications pay dearly for these convenient semantics.  Bit
errors result in relentless data retransmissions at the last hop, reducing the
efficiency of chains of upstream links that must wait for last-hop physical and
MAC-layer conditions to improve.  Constant error recovery also increases
endpoints' utilization of power-hungry wireless radios.

However, many applications deal with fundamentally noisy inputs and can tolerate
errors.  Computer vision, game mechanics, sensor data analysis, and search are
all in this category.  For these applications, low-layer integrity checking is
often superfluous; in fact, lower layers violate the End-to-End
Principle~\cite{end-to-end} by making judgments on the quality of network
traffic.  What is an error-tolerant application to do?

The solution to this mismatch is to put error handling under applications'
control.  If applications can claim possibly damaged data for themselves, rather
than delegating integrity enforcement to lower layers, the network stack can
provide better performance---crucially, lower latency and greater
throughput---for error-tolerant applications, and both sides of a network
connection can save precious energy on wireless radio use.  An ideal realization
of such a scheme would allow applications to transmit \emph{some} important
data---such as metadata---precisely while allowing errors in the rest.

% here's the meat and potatoes...
This paper explores the consequences of providing \emph{approximate} networking
semantics to general-purpose applications.
To this end, we propose, design, and evaluate \sysnamefull (\sysname), an
architecture for general-purpose approximate communication.  \sysname provides
configurable integrity checking at multiple network layers, allowing
applications to selectively compose traditional integrity-protected
communication and higher-throughput unprotected communication.
\sysname builds on UDP-Lite, a transport protocol designed to accelerate
multimedia streaming by relaxing transport-layer integrity
checks~\cite{Larzon99udplite}. \sysname adds support for lower-layer
integrity-check relaxation and a higher-level networking API for error-tolerant
applications.  \sysname requires minor changes to applications' code and a small
kernel patch under Linux, but it requires no hardware or physical-layer changes
and is otherwise backward compatible with existing networks.

% \xxx[lc]{It it is important to emphasize that the key contribution of this work is to extend the notion of approximation to the network stack and propose an API for that. SAP and the apps are just proofs of concept.}
\mypara{Contributions.}
This paper (1)
extends the concept of general-purpose \emph{approximate computing} to
\emph{selectively approximate networking}, in which error-tolerant applications can extend the benefits of approximation beyond individual
machines; and
(2) defines and evaluates \sysname, a cross-layer embodiment of selectively
approximate networking for off-the-shelf Wi-Fi devices.

Approximate network semantics offer several advantages for applications.  First,
an application that tolerates errors can increase its throughput by accepting
data more quickly; application-specific \emph{semantic} integrity checks can
replace checksums.
Second, under adverse network conditions, an application can avoid
unpredictable lag times due to retransmissions that disrupt application
quality.
Third, an application can achieve greater usable range~\cite{riemann:fec} or
tolerate more interference for a given output quality.
Finally, an application can save energy by reducing the number of bytes
transmitted.

Unlike past approaches to error-tolerant communications that have focused on a
single class of applications (e.g., streaming media~\cite{sen:sigcomm10,
softcast}), \sysname provides a \emph{general} mechanism that is format
agnostic and usable by any application.  The \sysname prototype is implemented
atop IP and 802.11, but its key properties are portable to
other stacks.

\section{Motivation and Background}\label{sec:motivation}

As wireless networks replace wired networks, transmission errors become
more burdensome.  Errors at the MAC layer trigger retransmissions that
decrease goodput and channel availability.  Errors become more pronounced as
transmission rates and distances increase.  These problems are generally not
present---or negligibly rare---in wired networks.

An unfortunate side effect of MAC-layer errors is that they have a
congestion-like effect on networks upstream.  To a sender several hops away,
MAC-layer retransmissions at the last hop manifest as delays in reaching the
receiver, which both sides incorrectly interpret as congestion, causing
retransmissions along the entire path.  Last-hop errors thus cause slowdowns
along paths that are \emph{mostly} reliable.

If slightly damaged payloads did not require retransmission, applications could
experience lower latency and greater goodput, and the reduction in wireless
delays could alleviate upstream congestion and reduce MAC-layer wireless
contention.  The key question is whether damaged payloads can still be useful to
applications.

\mypara{Characterizing wireless errors.}
A simple experiment in a noisy WiFi environment---described fully in
\secref{sec:eval}---yields a surprising result: even at high rates of frame
retransmission, the bit error rate (BER) in frames considered corrupt remains
below a threshold that would be tolerable to many applications.
Furthermore, BER does \emph{not} scale directly with the fraction of frames that
are retransmitted, suggesting that, on average, a small number of bad bits cause
a large fraction of retransmissions.

This paper proposes \sysnamefull (\sysname) as a response to retransmission
behavior that is inappropriate for error-tolerant applications. \sysname allows
applications to choose approximate or precise network semantics at run time to
suit their needs.  Error-tolerant applications like those studied in previous
literature on approximate computing (e.g., image rendering, gameplay updates)
can choose \emph{approximate} transmission for information that is noisy or
contains uncertainty.  Applications that already reduce resource use via
approximate computing can match their communication strategies to their
computation strategies.
The main goal of \sysname is to reduce the time and energy that applications
must spend on communication.

\subsection{Background: Data Integrity}\label{sec:background:net}
The OSI layered networking model defines a stack of abstraction layers that
enable progressively simpler communication primitives.  The bottom layer
encapsulates the physical properties of the network, e.g., radio transmission
encoding, and the top layer encapsulates applications' communication via
straightforward \lil{send} and \lil{receive} operations.
This model delegates responsibility for well-formedness to each layer.  This
section focuses on 802.11 (Wi-Fi) networks running IPv4, but many of the terms
are applicable to other wired and wireless networks, mutatis mutandis.

In 802.11 networks, \term{stations} transmit \emph{frames} via a decentralized
Media Access Control (MAC) protocol that coordinates access to the wireless
channel.  Each frame includes a mandatory 32-bit \term{frame check sequence}
(FCS) that is a cyclic redundancy check (CRC) of the rest of the frame.
The standard does not mandate a specific action for stations to take when an FCS
check fails; instead, it says that ``All STAs shall \emph{be able} to validate
every received frame using the frame check sequence''~\cite[\S8.1]{802.11-2012}
(emphasis added) and it mandates that stations will issue \term{positive
acknowledgments} (ACKs) upon correctly receiving certain types of
frames~\cite[\S9.3.2.2]{802.11-2012}.  Absent an ACK, senders retransmit frames
up to a retry limit.

At the network layer, IPv4 packets carried in 802.11 frames include another
16-bit
checksum that covers only the IP header~\cite{rfc791}.  The IPv4 standard mandates that receivers ``silently
discard every datagram that has a bad checksum.''~\cite[\S3.1]{rfc791}

At the transport layer, TCP and UDP include packet (datagram)
checksums~\cite{rfc1122}.  The 16-bit TCP checksum is required for TCP packets;
receivers must silently discard packets that fail the CRC check.  UDP, however,
allows senders to fill a datagram header's 16-bit checksum field with zero bits,
which tells the receiver not to compute a checksum over the datagram, and
which may save checksum-computation time for the sender and receiver.

The net effect of these integrity checks across layers is that \emph{multiple
independent mechanisms can cause damaged data to be discarded}.  If the link
layer did not include a checksum, then transport-layer checks would be the only
line of defense against corruption.  In modern networks that include FCS-like
low-level integrity checks (e.g., Ethernet and 802.11), higher-level checks are
largely redundant.  In response, IPv6 omits header checksums because they are
redundant with lower- and higher-layer checksums meant to protect against bit
errors and format problems~\cite{rfc2460}.

For the protection they provide, checksums are not perfect; they are inadequate
on their own to provide perfect data integrity.
\iffalse
Modern software network stacks, including Linux's \lil{mac80211} subsystem,
support \term{checksum offloading} that allows hardware to compute checksums and
signal success or failure in special descriptors.  Checksum offloading makes the
performance cost of checksums negligible.
\fi
Checksums \emph{do} offer receivers some protection against malformed data from
ill-behaved network stacks, but the simple CRCs on network packets offer
imperfect, skewed output distributions and cannot detect certain kinds of
splices and other small changes~\cite{stone:crcs}.
Consequently, the traditional advice to application designers is to perform
application-level checks once data arrives.  Application-level integrity checks
thus represent \emph{another} layer of protection---one that is arguably more
important than the lower-level checks because it can incorporate arbitrary
decision-making logic and quality control.  With respect to the
End-to-End Principle~\cite{end-to-end}, the application-layer check is the most
important.

\mypara{Relaxing integrity checks.}
The UDP-Lite variant of UDP~\cite{Larzon99udplite, rfc3828} exists to provide
\emph{configurable} checksum protection of UDP packets for applications that can
tolerate some errors.  UDP-Lite replaces the UDP header's \emph{Length} field
(which can be inferred from the enclosing IP datagram's \emph{Length} field)
with a \emph{Checksum Coverage} (\lil{cscov}) field that specifies the number of
octets in the UDP-Lite datagram to be included in a checksum computation.  A
value of $0$ in the checksum coverage field indicates full coverage, equivalent
to a UDP checksum.  The UDP-Lite header is always included in the checksum
computation.  As in UDP, datagrams with errors in the \lil{cscov}-covered range
are unceremoniously dropped.

Applications can control UDP-Lite checksum coverage via \lil{setsockopt} calls
at run time. An application can provide a \lil{cscov} value so that none, all,
or the first $n$ bytes of a payload are covered by the
checksum.

Past work has shown \emph{in simulation} that UDP-Lite can decrease loss rates
for multimedia applications by deeming partially damaged data
acceptable~\cite{hammer:speech, meriaux:udplite}.  Like link-layer coding
techniques that implement packet and header correction from redundant
information~\cite{marin:mac-lite}, these techniques' tolerance of damaged
payloads can mostly preserve application-layer quality metrics while reducing
retransmission rates.

\sysname uses UDP-Lite to provide configurable \emph{approximate} semantics to
general applications beyond multimedia, offering approximate communication as a
complementary mode to \emph{precise} guaranteed in-order delivery.  \sysname
combines these transport-layer mechanisms with the MAC-layer mechanisms
necessary to preserve the semantics across the network stack, embracing the
end-to-end principle~\cite{end-to-end} to allow applications on both ends of a
communication link to fully control data delivery.

\subsection{Background: Approximate Computing}\label{sec:background:approx}

The growing body of work in approximate computing observes that many
applications can save time and energy by engaging fewer resources---whether by
skipping work~\cite{hoffmann:knobs}, reducing voltage~\cite{truffle-asplos12},
storing information in less reliable memories~\cite{flikker,
approxstorage-micro13}, or training neural networks to emulate
expensive computation~\cite{npu-micro12}.

Applications that benefit from approximate computing include 3D rendering and
game engines~\cite{enerj-pldi11}, database query processing~\cite{blinkdb,
approx-queryproc}, computer vision and robotics~\cite{npu-micro12}, sensor data
storage and analysis~\cite{approxstorage-micro13}, and weather
simulations~\cite{weatherpalem}.

To date, approximate computing's performance and efficiency improvements have
largely been confined to \emph{local} speedups on a single device.
Unfortunately, energy-constrained devices that benefit from these speedups, such
as phones or embedded devices, can
waste all of these performance gains on data transmission---even when the data's
recipients do not require perfect fidelity.  Reliable TCP/IP stacks on reliable
MAC layers are imperfectly matched with these application domains.

Previous work on approximate computing has not explicitly sought to optimize
\emph{communication} costs, leaving an opportunity for further savings.  Radios
are major consumers of power in modern computers.  On a mobile phone, the WiFi
and cellular radios require an order of magnitude more power than even the CPU
or memory~\cite{Carroll_Heiser_10}.  Further, transmitting data requires not
only the radio but the CPU to be awake, since network stacks are predominantly
implemented in software.

\begin{figure}[t]
\begin{center}
\includegraphics[width=0.42\columnwidth]{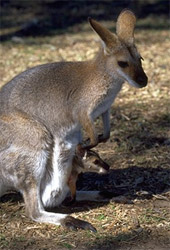}
\includegraphics[width=0.42\columnwidth]{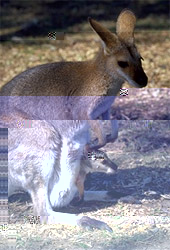}
\end{center}
\caption{A JPEG image transmitted in \sysname's precise (left) and approximate
(right) modes showing that bit errors render the image degraded but
recognizable.  \footnotesize{(Photo of a red-necked wallaby [\textit{Macropus
rufogriseus}], \copyright\,2002 California Academy of
Sciences.)}}
\label{fig:wallaby}
\end{figure}

Even heavily encoded formats, such as JPEG, are often robust against a certain
amount of error (\figref{fig:wallaby})~\cite{frescura:jpeg2000}.  The intuition
behind this surprising robustness is that the information that influences human
perception---such as low-frequency image components---often comprise only a
small part of the total image data, with potentially inconsequential details
(e.g., high-frequency components or color tables) comprising most of the bits.
On one hand, a randomly placed bit error is more likely to affect
inconsequential data; on the other hand, unfortunately located bit errors can
completely hamper decodability.
The necessary division of data between precise and approximate handling,
proposed for EnerJ~\cite{enerj-pldi11}, can also apply to data formats being
transmitted: precise data can receive checksum coverage, and approximate data
can be transmitted without checksum coverage.

Of course, certain kinds of data are fundamentally precise and are incompatible
with approximate transmission.  Encrypted or heavily compressed payloads are
prime examples of such precise data.  Still, a great deal of Internet and LAN
traffic occurs without encryption or compression~\cite{sandvine-report}, so
mechanisms to relax integrity requirements still apply.

\section{Design of \sysname}\label{sec:design}

\sysname is a transport protocol and API designed to expose potentially damaged
data to applications that want it.
\sysname's cross-layer design philosophy can be summarized as
follows:
\begin{compactitem}
  \item Integrity checking should be under \emph{applications'} control;
  applications should be able to receive potentially damaged data if they wish.
  \item Applications should be allowed to switch between precise and approximate
  communication modes quickly and without special privileges.
  \item For reasons of cost and inertia, approximate communication should be
  possible on unmodified hardware and physical layers and should travel
  concurrently with conventional traffic.
\end{compactitem}

This design philosophy underlies the \term{approximate networking model}, which
we define as follows:
\begin{compactenum}
  \item A unit of data is either \emph{precise} or \emph{approximate}.
  \item Precise data will have accompanying metadata (e.g., a checksum) that the
  receiver can use to confirm exact reception.
  \item Approximate data can optionally include similar metadata, but
  sub--application-layer mechanisms at the receiver will not use this metadata
  to decide whether to pass the data upward.
  \item A single transmission can include both precise and approximate data,
  provided it includes metadata that unambiguously marks the two kinds.
  \item Precise data requires acknowledgments from receivers and is guaranteed
  to arrive in order; approximate data does not.
  \item Control data (e.g., acknowledgments, connection setup, and teardown) are
  precise.
\end{compactenum}

\sysname embodies the approximate networking model by building it atop
802.11~\cite{802.11-2012} and UDP-Lite~\cite{Larzon99udplite}.  It necessarily
involves multiple layers for the reasons outlined in
\secref{sec:background:net}, and maps to the model as follows:
\begin{compactenum}
  \item \sysname provides sockets that are precise by default, and an API to
  switch the socket to approximate mode and back.
  \item Sockets are built on UDP-Lite datagrams that offer partial or full
  checksum coverage of precise data.
  \item \sysname adds an optional \emph{approximate} mode to 802.11 to carry
  approximate data, selectively changing retry behavior.
  \item A checksum coverage field inherited from UDP-Lite demarcates where
  precise data ends and approximate data begins within a datagram.
  \item Receivers send (precise) acknowledgments for precise data; senders
  re-send precise data until they receive acknowledgments.
  \item \sysname always uses precise transmissions for control data including
  connection setup and teardown; it implements a simple TCP-like handshake.
\end{compactenum}

The remainder of this section details the design of \sysname with reference to
conventional wireless network communication using 802.11 and TCP/IP.
\secref{sec:implementation} gives lower-level implementation details.

\subsection{Changes to 802.11}
\sysname makes only one small software change to 802.11 transmission: the 802.11
driver examines outgoing payloads---specifically, inspecting \lil{sk_buff}
structures for UDP-Lite headers that indicate partial checksum coverage---to
detect whether the application has marked the data as approximate.  If so, the
driver sets the 802.11 retry limit to zero to disable retransmission of the
frame if it does not receive an acknowledgment frame.

Receivers are subject to a larger number of software changes to support
approximate data.
\sysname provides a boolean driver-level switch that controls both hardware
checksum offloading (disabled when the switch is in approximate mode) and the
kernel's decision procedure that leads to each frame's acceptance or rejection.
When the switch is in the \emph{approximate} position, the 802.11 driver
processes \emph{all} frames instead of simply dropping those that fail the FCS
check.  The MAC-layer logic is left intact, so the driver will, for example,
still ignore frames destined for other stations (which may include frames with
corrupted address fields).

With the driver-level mode switch in the \emph{precise} position, 802.11
behavior is unchanged from conventional 802.11 transmission and reception.

\subsection{Changes to Applications}

Because it offers its own socket API, \sysname requires changes to application
code.  The API presents \term{send} and \term{receive} functions that map
exactly onto the familiar sending and receiving functions of conventional
sockets.  The behavior of these functions depends on the socket's
\emph{approximation mode}, which is \emph{precise} at socket creation but can
switch between \emph{precise} and \emph{approximate} at run time.  The socket's
approximation mode is independent of the aforementioned kernel-level switch.

Applications that use UDP, such as media-streaming applications, are trivially
portable to \sysname, especially if they do not rely on guaranteed in-order
delivery.  Since \sysname sockets wrap UDP sockets, an application that already
uses datagram-oriented communication need only substitute calls to the \sysname
versions of the socket functions, and optionally change the approximation mode
based on application requirements.

Porting TCP applications is only slightly more challenging.  As a
proof-of-concept embodiment of the approximate networking model, \sysname
reimplements only the most important property of TCP on which applications rely:
guaranteed in-order delivery for precise data.  It does so via the same
mechanism that TCP uses, namely ACKs, though it sends them as separate
(precise) transmissions rather than header fields.
The more significant difference is that \sysname, based on UDP, is datagram
oriented rather than stream oriented; application code must be more explicit
about when to send and receive data.  The same problem affects any port of a
TCP-based application to UDP, so we do not belabor it here.  \secref{sec:eval}
describes porting a TCP-based application to \sysname.

Regardless of any changes to socket code, applications using \sysname become
responsible for data integrity.  In contrast to cheap but imperfect checksums,
applications can implement semantic, application-aware integrity checks (e.g.,
``is this data point within $n$ units of the previous data point?'') when they
accept approximate, potentially damaged data.

\subsection{Design Implications}

According to the standard, an 802.11 transmitter is allowed to adjust its
bitrate, maximum transfer unit (MTU), selected channel, transmit power, retry
limit, request-to-send (RTS) threshold, and power-management options.
Retransmission rate, as a proxy for channel quality, is one metric that governs
how aggressively a transmitter adjusts these parameters.  Each of these causes
performance to degrade under poor channel conditions, so stations often
overprovision for current conditions.  \sysname's introduction of error
tolerance enables stations to adjust these parameters more aggressively,
reducing overprovisioning.

Under \sysname, an 802.11 station can: reduce its transmit power (which makes
decoding errors more likely by reducing the signal-to-noise ratio); increase its
bitrate (which makes decoding errors more likely by making symbols ``smaller'');
increase its RTS threshold (which makes bit errors in frames more likely by
gambling that longer transmissions will be error free); and decrease its retry
limit as described above (since each try is more likely to succeed).
At the MAC layer, the operating system can increase the maximum transfer unit
(MTU), effectively gambling on longer transmissions in the same way as
increasing the 802.11 driver's RTS threshold.

\subsection{Implementation}\label{sec:implementation}

\sysname comprises two parts: a small set of device driver
and software MAC layer modifications and a user-space library that implements
\sysname's user-space protocol components.
We implemented the kernel modifications against version 3.8.13 of
the Linux kernel.

Our changes to the Linux kernel are: 40 additional lines of code in the
\lil{mac80211} subsystem, which implements the 802.11 MAC layer, to modify its
frame-dropping logic; and 165 new lines of code in the \lil{ath5k} 802.11 card
driver, to modify its frame handling and implement \sysname's approximation mode
switch.

The \sysname user-space library, \lil{libsap}, implements an API atop UDP-Lite
sockets and comprises 766 lines of C code.  A \lil{sap_sock_t} socket object
keeps metadata to track acknowledgment numbers for precise datagrams and the
socket's connection state; there is no global state.  An application can act as
sender, receiver, or both with the appropriate API calls.

For senders, three API functions handle connection setup and teardown.  The
\lil{sap_connect} function stores endpoint information in a socket object
so it is retained between datagrams.  \lil{sap_connect} then sends a precise
\lil{SAP_PING} message to the receiver and waits for an acknowledgment.  (If no
acknowledgment arrives within a timeout period, the \lil{sap_connect} call
returns an error; it is ideal to set up the receiver to listen before the sender
issues a ping.)  After the receiver acknowledges the ping, the socket is in the
connected state and the sender is free to send datagrams.
When it is done with a \sysname transmission (e.g., an approximate file
transfer), the sender closes its end of the socket by issuing a precise
\lil{SAP_FIN} message to the receiver.  The sender waits up to a timeout period
for the receiver to acknowledge the \lil{SAP_FIN} message, then calls
\lil{sap_close} which simply frees the socket's memory.  (It is wise for senders
to use multiple threads of execution so that precise transmissions, such as
\lil{SAP_FIN}, do not block traffic for other connections.)

Receivers use \lil{sap_listen} to set up a \sysname socket, which implements the
other side of the above protocol on top of UDP-Lite.  For brevity, we omit
the details of the receive side.  The salient property of receivers is that they
handle precise and approximate traffic differently.

\iffalse % sadly no longer relevant to this draft (no TCP yet)
\mypara{802.11 device driver and \lil{mac80211} subsystem changes.}
\sysname adds a new frame subtype, \lil{ST_APPROX_DATA}, to each 802.11 frame that
should be transmitted approximately.  To simplify the API for applications,
\sysname determines automatically whether to set the frame subtype by looking
for \sysname custom fields in the headers of transport-layer packets.

\mypara{TCP stack changes.}
\sysname adds a custom TCP socket option \lil{TCP_SAP_APPROX}, which Listing
\ref{lst:sockopt} shows how to use.  The socket option carries a 32-bit
\lil{length} value that specifies how much of the packet's prefix should be
covered by a checksum.  This specifier is modeled after the partial-checksum
specifier of UDP-Lite~\cite{rfc3828}.

\mypara{User-space socket library.}
Applications based on UDP use \sysname's user-space socket library,
\lil{libsap}, to send and receive UDP-Lite datagrams instead of plain UDP
datagrams; this library must be linked into the application.
\fi % TCP

\section{Evaluation}\label{sec:eval}

This section evaluates the degree to which \sysname meets its design goals---in
short, how approximate transmission of data affects applications'
performance--quality balance.

The evaluation studies three applications, each time focusing on a different
aspect of \sysname's behavior:
a simple synthetic application that streams precomputed values to a receiver
over an approximate socket;
a web server (lighttpd) that streams files to clients in precise or approximate
mode; and
a location-tracking application that emits periodic GPS readings for speed
calculation.

\iffalse
The remainder of this section describes experiments that:
\begin{compactitem}
\item Characterize 802.11 transmission errors.
\item Test the effect of range on approximate communication; vary the location
of the transmitter.
\item Simulate different packet loss rates?  Other pathological channel
conditions?
\item Compare \sysname{} to other systems for approximate communication, e.g.\
Sen's~\cite{sen:sigcomm10}.
\item Compare simulated or real energy consumption.  Can we effectively monitor
power consumption of a wifi card?  Investigate a PCI shim instrumented with a
sense resistor or Hall effect sensor to measure current draw at fine
granularity.
\end{compactitem}
\fi

We evaluated \sysname between PCs in a multistory office environment with
interference from many nearby enterprise-grade access points.
For \sysname senders and receivers, we used Dell Optiplex GX620 workstations
running Ubuntu Linux 12.04.5 LTS and kernel version 3.8.13 (patched as described
in \secref{sec:implementation}).  Each workstation used a TP-Link WN350GD
802.11b/g PCI Express card controlled by the \texttt{ath5k} kernel driver.
When testing transmission between two nodes using \sysname, we kept one node
stationary in a $3\,\text{m} \times 4\,\text{m}$ office and moved the other node
to different points in the building, as close as $1\,\text{m}$ and as far away
as $12.2\,\text{m}$.

\subsection{Applications}\label{sec:apps}

\mypara{Streamer.}
We designed the first application, \emph{Streamer}, to characterize 802.11 frame
errors and their effect on application-level data delivery.
Streamer sends a fixed number of datagrams to a receiver exclusively in
approximate mode.  The receiver captures and analyzes all data that reaches the
receiving application, and it also collects and analyzes 802.11 statistics from
the Linux kernel.

The sender and receiver agree on the contents of the entire stream in advance by
choosing an initial 32-bit seed, which the sender sends to the receiver
precisely, and which is incremented for each datagram.  Each datagram consists
of the result of passing the 32-bit counter through a hash function that outputs
32 unbiased bits.

\mypara{Lighttpd.}
We chose a web server application to evaluate (1) the difficulty of porting a
TCP-based application to \sysname and (2) the throughput performance of
\sysname's precise and approximate modes versus precise TCP transmission.
\emph{Lighttpd}~\cite{lighttpd} is a popular event-based web server.  Web
servers are especially good candidates for \sysname's \emph{selectively}
approximate communication because they serve a variety of file types via the
ubiquitous HTTP protocol.  A typical webpage visit from a browser loads several
resources---HTML pages, scripts, and images---often from the same server, but
only some of these data types are damage tolerant.
We modified lighttpd to serve content via \sysname as follows.

HTTP requests proceed as normal, with clients initiating connections to the
web server via TCP and issuing requests.  A \sysname-aware client adds two
headers: \lil{X-SAP-Approx} followed by a list of MIME types for which the
client can tolerate errors (e.g., \lil{image/jpeg}); and \lil{X-SAP-Port}, which
tells the server on which UDP port the client will listen for a \sysname
connection.

Upon receiving an HTTP request with \sysname headers, the modified lighttpd
loads files from backing storage and determines their MIME types as usual, but
it also matches MIME types against those that the client requested to receive
approximately.  In the case of a match, the server initiates a \sysname
connection back to the client on the client's requested port, sets the \sysname
socket's mode to \emph{approximate}, and sends the contents of the file in
$1\,\text{KB}$ datagrams.  Files whose MIME types do not match the client's
\lil{X-SAP-Approx} header are sent via TCP.

For a direct comparison of precise and approximate transmission, we also added
an \lil{X-SAP-Force-Precise} HTTP request header to tell lighttpd to send its
responses via \sysname in precise mode instead of approximate mode.

A set of concurrent HTTP requests may cause the server to initiate multiple
\sysname connections to the client, but for simplicity, \sysname includes no
multiplexing facility.
To regain the parallelism that multiple TCP streams offers, clients can issue
each HTTP request with a different \lil{X-SAP-Port} value, then \lil{sap_listen}
on all of the corresponding ports at once.

These changes added 178 lines of C code to lighttpd version 1.4's 54632 lines
and required only a few hours of work for a single programmer, most of which
time went to understanding lighttpd's request workflow.

\mypara{Tracker.}
To evaluate \sysname's effect on the \emph{quality} of transmitted data, we
built a location-tracking application with a straightforward quality metric.  In
practice, quality metrics are application specific.  Applications that use
approximate communication with \sysname should implement their own quality
metrics that perform sanity checking and damage control (e.g., requesting
retransmission of unacceptably damaged pieces of data).

\emph{Tracker} is a simple location tracker that emits a latitude and longitude
(a pair of 64-bit double-precision floating point values) at a predefined rate.
A sender emits these values one at a time and a receiver calculates the
cumulative moving average of the sensor's speed with each update, finally
emitting a final moving average covering the duration of the trace.  We used a
trace of GPS readings collected by a mobile phone while its owner walked around
a campus.  To smooth the inevitably noisy GPS signal, we followed the example of
previous approximate-computing work~\cite{uncertaint} and made the receiver
discard location points that would have indicated foot speed outside a
reasonable bound---in this case, any speed greater than the maximum running
speed of an Olympic sprinting medalist.

\iffalse
Since Tracker's quality metric is cumulative and history dependent, the sender
transmits the first three data readings (48 bits total) precisely before
switching to approximate mode; this ensures that the moving average starts at a
reasonable value.
\fi

In a real deployment of this application, a moving node associated with a WiFi
network (via one or many access points) would experience highly variable channel
quality~\cite{vifi}, making continuous reliable transmission over TCP difficult.
The \sysname version of Tracker aims to constrain tracking-data error metrics
while providing steady transmission that is not fraught with delays.  We also
implemented a TCP version that operates analogously.

The \sysname version of Tracker comprises 152 lines of C for the listener, 101
lines of C for the sender, and 34 lines of shared C code for dealing with GPS
measurements.  The TCP version comprises an extra 50 lines of code, mostly for
buffer management.

\subsection{Characterizing Frame Retransmission Behavior}

We used the \term{Streamer} application to transmit predictable, unbiased bit
patterns in order to understand two factors influencing \sysname's performance:
the proportion and nature of retransmitted frames, and how well \sysname might
be able to recover usable information for receivers that can tolerate damage.
All tests were performed during the daytime in a busy office building with many
enterprise access points nearby.  When we varied the distance between nodes,
distances over $5\,\text{m}$ meant the moving node was in nearby rooms on the
same floor (i.e., transmissions traversed walls).  We set up an ad hoc
802.11 network with an invisible SSID for \sysname nodes.

\begin{figure}[t]
\centering
\includegraphics[width=0.8\columnwidth]{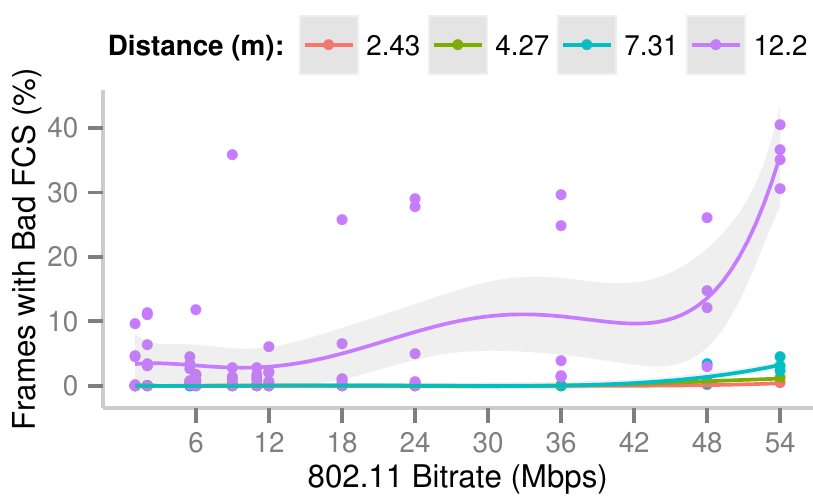}
\caption{The fraction of frames that must be retransmitted increases
dramatically with increasing 802.11b/g bitrate and distance.}
\label{fig:frame-corruption}
\end{figure}

To measure rates of frame retransmission in an 802.11 network, we sent
$10\,\text{MB}$ of predictable data between two of our test machines using the
Streamer application, varying the location of one of the nodes and testing five
times per location ($50\,\text{MB}$ total).  We used \sysname in \emph{precise}
mode so that the sender would retransmit frames.  \figref{fig:frame-corruption}
plots the fraction of frames that the sender retransmitted at least once because
the receiver, seeing they did not pass the FCS check, did not acknowledge them.
The salient feature of \figref{fig:frame-corruption} is that
distance and bitrate have a deleterious effect on correct frame transmission,
with retransmission rates exceeding $30\%$ on average when the nodes used the
highest bitrate at $12.2\,\text{m}$.

\begin{figure}[t]
\centering
\includegraphics[width=0.8\columnwidth]{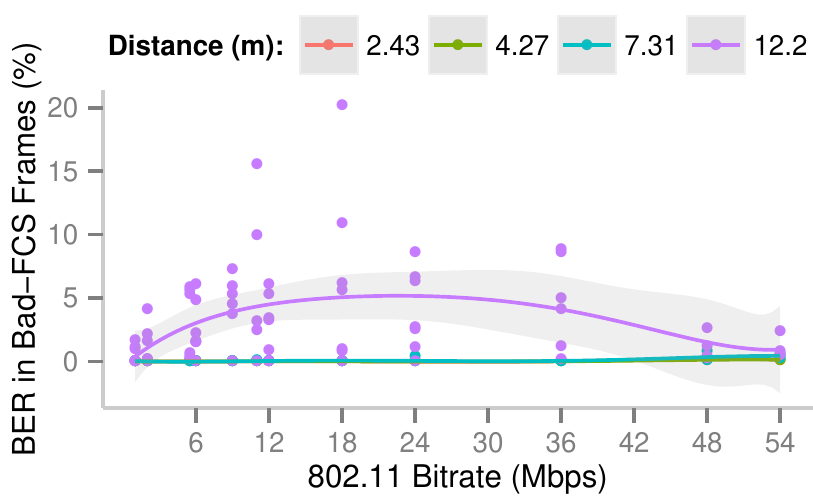}
\caption{Bit error rate does not scale directly with the fraction of corrupt
frames.  At higher bitrates, despite a greater fraction of frames being corrupt,
coding gains result in decreasing BER in damaged frames.}
\label{fig:low-ber}
\end{figure}

A second question is: if frames were to be accepted despite FCS failures, how
bad would the damage be, from an application's perspective?
To answer this question, we sent $10\,\text{MB}$ in \sysname's
\emph{approximate} mode via the Streamer application and measured the bit error
rate (BER) of the payloads that arrived in damaged frames.  We varied the
distance between the sender and receiver and performed five trials at each
distance as in the previous experiment.
\figref{fig:low-ber} plots BER versus bitrate and distance for this experiment.
Even at distances and bitrates with a high proportion of damaged frames, the bit
error rate remained low.  Most notably, at $48$ and $54\,\text{Mbps}$, coding
gains drive the BER well below $5\%$ in damaged frames.  Counting the
non-erroneous bits in damaged frames along with all the bits in correctly
received frames, transmissions at the highest bitrates consisted of $99.6\%$
correct bits---an error rate that is tolerable to many
applications~\cite{enerj-pldi11, blinkdb}.
\xxx[br]{Find a ``best'' citation for this level of error being acceptable.}

\begin{figure}
\centering
\includegraphics[width=0.8\columnwidth]{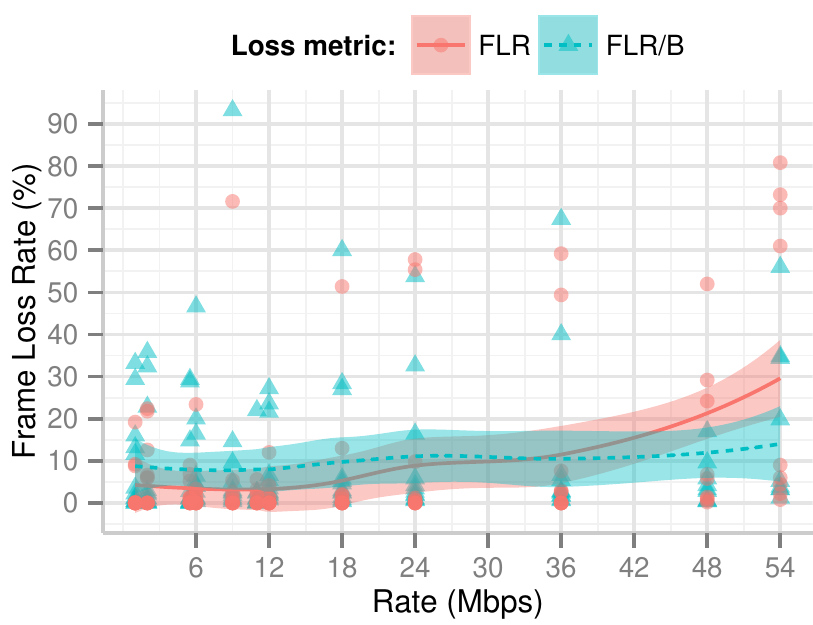}
\caption{Accepting damaged frames (FLR/B, dotted blue line) with \sysname allows
a receiver to decrease the effective frame loss rate (FLR, solid red line).}
\label{fig:plr-vs-bitrate}
\end{figure}

Finally, to forecast the effect of accepting damaged data, we measured the frame
loss rate (FLR) in the Streamer application several ways.  At all the positions
described in the previous experiment, we transmitted
$10\,\text{MB}$ in \sysname's approximate mode and measured: (1) the fraction of
frames that did not arrive at the receiver, and (2) the fraction of frames that
arrived with damage (i.e., that failed their FCS checks).  Without \sysname's
modifications to the Linux kernel, the effective frame loss rate would have been
the sum of the missing and damaged frame rates.  With \sysname's modifications,
the effective frame loss rate reflects only the fraction of \emph{lost} frames
(which we call FLR/B), since \sysname and the Streamer application make no
attempt to recover these.
\figref{fig:plr-vs-bitrate} shows the rates of payload corruption and loss over
all bitrates and distances.
Note that packet loss may \emph{also} be acceptable to applications that can
accept damaged data, suggesting that a nonzero frame loss rate is reasonable.

\subsection{Web Throughput}\label{sec:eval:faster-web}

To evaluate \sysname's effect on throughput for a complete application
(lighttpd), we repeatedly sent large compressed (JPEG) images from a sender to a
receiver, both nodes on an ad-hoc network of the machines described earlier.
Nodes were connected to the building network via wired Ethernet, which we used
as a backchannel to coordinate the sender and receiver's configurations between
experiments.  Lighttpd listened on both the Ethernet and 802.11
interfaces.

\mypara{Throughput versus channel quality.}
We used the distance between sender and receiver as a proxy for channel quality,
and measured how long it took to transfer an identical file in several different
ways: via (precise) TCP, precise \sysname, and approximate \sysname.

In this experiment, the receiver first warmed the sender's cache by requesting
the file over TCP via the wired Ethernet interface.
The receiver then issued sequential HTTP requests via the wireless interface,
with one-second pauses in between the completion of a request and the beginning
of the subsequent one.  The receiver issued 100 requests over TCP without
\lil{X-SAP-*} headers to request that the HTTP response be sent on the same TCP
socket; then 100 requests over TCP for the same file with an
\lil{X-SAP-Force-Precise} header to send the file back via \sysname in precise
mode; then 100 requests over TCP for the same file to be sent back via \sysname
in approximate mode.
We fixed the bitrate at $54\,\text{Mbps}$, the maximum for the 802.11g cards,
and measured the total time from the end of the receiver's HTTP request to the
sender's precise \lil{SAP_FIN} packet indicating the end of the transmission.

\begin{figure*}
\centering
\includegraphics[width=0.9\textwidth]{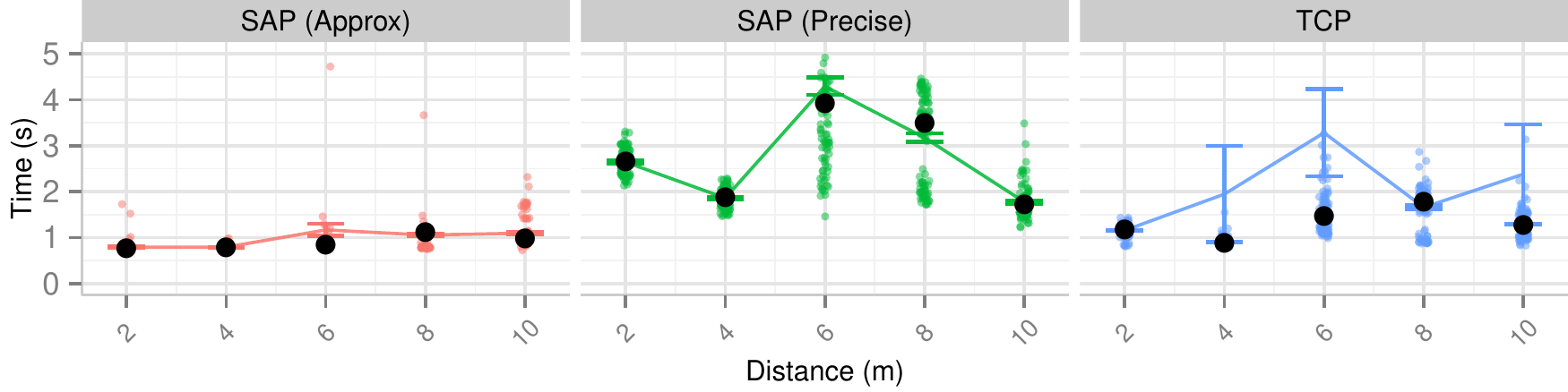}
\caption{At a fixed 802.11 bitrate of $54\,\text{Mbps}$ and increasingly with
distance, \sysname in approximate mode transfers a $2\,\text{MB}$ file more
quickly than TCP or precise \sysname.  Black dots represent median transfer
times over 100 trials; error bars are centered around mean transfer times and
represent standard error.}
\label{fig:precise_approx_Ndistance}
\end{figure*}

\begin{table}[h]
\centering
\begin{tabular}{c|c|c|c}
& TCP & \sysname (Approx) & \sysname \\
Distance (m) & Time (s) & Time (s) & Speedup \\
\hline
2 & 1.163 & 0.765 & $1.52\times$ \\
4 & 0.878 & 0.785 & $1.11\times$ \\
6 & 1.452 & 0.828 & $1.75\times$ \\
8 & 1.769 & 1.118 & $1.58\times$ \\
10 & 1.260 & 0.968 & $1.30\times$ \\
\hline
\multicolumn{3}{r}{\em Geom.\ mean\hspace{1em}}& $1.43\times$
\end{tabular}
\caption{Speedup of \sysname in approximate mode compared to TCP for a
$2\,\text{MB}$ transfer.}
\label{tbl:www-speedup}
\end{table}

\tblref{tbl:www-speedup} and \figref{fig:precise_approx_Ndistance} show the
results of the above experiment with a $2\,\text{MB}$ randomized binary file
over TCP and both modes of \sysname.  In this experiment, \sysname in
approximate mode achieved the shortest transfer time at all distances, with an
average speedup of $1.43\times$ versus TCP.
TCP was the second-fastest transport layer, roughly twice as slow as \sysname in
approximate mode.  At several distances, TCP transfer time increased by
nearly two orders of magnitude, skewing the mean transfer time---an unacceptable
failure mode for applications that require low inter-packet latency.
For this application, \sysname's precise mode used a fixed
payload size of 1~KB and did not attempt to implement any form of payload size
adjustment, backoff, congestion control, or ACK coalescing, so its slowness
relative to the well-tuned Linux TCP stack is not particularly surprising.

\mypara{Throughput versus bitrate.}
For this experiment, we fixed the distance between two nodes at $4\,\text{m}$
(in the same room), then varied the bitrate across the full range of 802.11b/g
bitrates.  The sender sent an identical $1\,\text{MB}$ file via the wireless
interface using each of the three protocols from the previous experiment, ten
times per protocol.
As before, we sent two initial requests via the wired Ethernet interface to warm
the sender's lighttpd cache.
We then repeated the experiment at $6\,\text{m}$ and $8\,\text{m}$.

\begin{figure*}[t]
\centering
\includegraphics[width=0.9\textwidth]{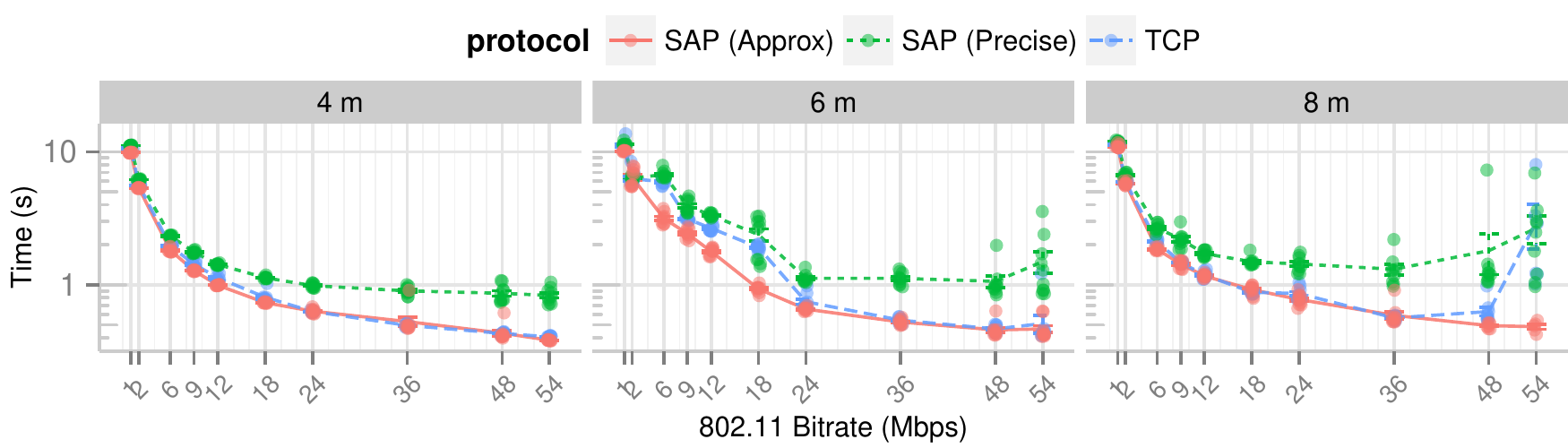}
\caption{At a fixed distance of $4\,\text{m}$ offering good channel conditions,
\sysname's throughput tracks TCP's.  As conditions worsen (at $6\,\text{m}$ and
$8\,\text{m}$), \sysname's throughput exceeds TCP's at nearly every bitrate,
with less variance.  Error bars are centered at the mean over 10 trials for each
$\langle \text{bitrate},\text{distance},\text{protocol}\rangle$ setting.}
\label{fig:tput-bitrate}
\end{figure*}

\figref{fig:tput-bitrate} shows the results of the variable-bitrate experiments.
When channel quality is good, as it is at $4\,\text{m}$, retransmissions impose
little overhead on precise transmissions, so \sysname's savings are not
significant.
As distance increases, channel quality degrades accordingly, and the precise
protocols' transfer times increase at higher bitrates because damaged frames
require retransmission.
A worse problem from many applications' perspective is that the
variance of transfer times increases sharply as channel quality drops.  \sysname
in approximate mode exhibits less variance than either precise protocol.

The plots suggest several potential strategies for applications that are faced
with falling throughput.  Keeping TCP or precise UDP but switching to a lower
802.11 bitrate is one option, since longer symbols on the air and different
coding schemes lessen the impact of interference.  However, 802.11 bitrate
selection is typically not exposed to applications.
An application using \sysname can respond in another way, by decreasing the
amount of protection applied to payloads.  According to these experiments, this
alternative approach should allow the application to retain reasonable
throughput despite channel degradation.

\subsection{Result Quality}\label{sec:eval:quality}
To measure the effect of approximate communication on an application with a
quality metric, we used the \emph{Tracker} application to compute the cumulative
moving average of a walking person's variable speed, using a recorded trace of
945 data points.  The ground-truth average speed, computed directly on the
original trace, was $1.56\,\text{m/s}$. The cumulative moving average smooths
inevitable noise from the GPS receiver and also helps the receiving application
smooth incoming readings that arrive with damage.

We set up two \sysname nodes as in previous experiments and designated the fixed
node as the receiver and the moving node as the transmitter.  We fixed the
bitrate at $54\,\text{Mbps}$ and varied the distance between the sender and
receiver from $2\,\text{m}$ to $15\,\text{m}$.  At each distance, we collected
20 data points with both TCP and \sysname in approximate mode, computing the
average walking speed.

\begin{figure}[t]
\centering
\includegraphics[width=0.8\columnwidth]{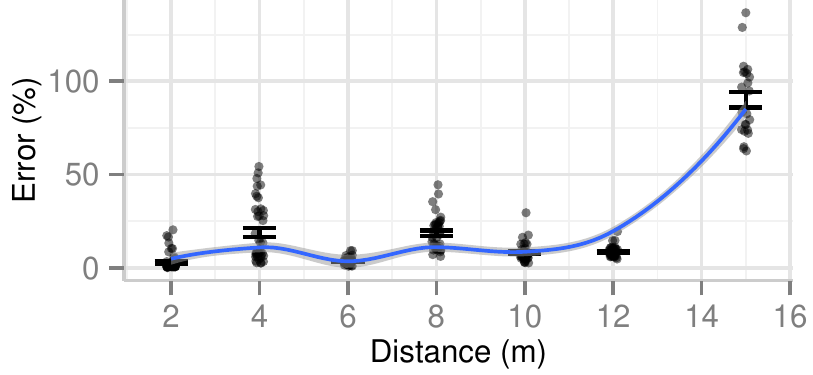}
\caption{For a location-tracking application, varying channel quality (via
distance, x-axis) with a fixed bitrate causes output quality (cumulative moving
average over a trace, ground truth $= 1.56\,\text{m/s}$) to degrade gracefully
until over half of the values are lost (at $15\,\text{m}$). Error bars
are centered around the mean and represent standard error over $50$ trials per
distance.}
\label{fig:quality-distance}
\end{figure}

\figref{fig:quality-distance} summarizes the quality degradation we found in
this experiment.  The median error at each distance was less than
$10\%$ versus the ground truth, except at $8\,\text{m}$ and $15\,\text{m}$,
where the $18.0\%$ and $83.4\%$ median error respectively represent
$0.28\,\text{m/s}$ and $1.40\,\text{m/s}$ error in the final speed calculation.
At these distances, over half of the values were lost or discarded because of
frame errors.

\begin{figure}[t]
\centering
\includegraphics[width=0.8\columnwidth]{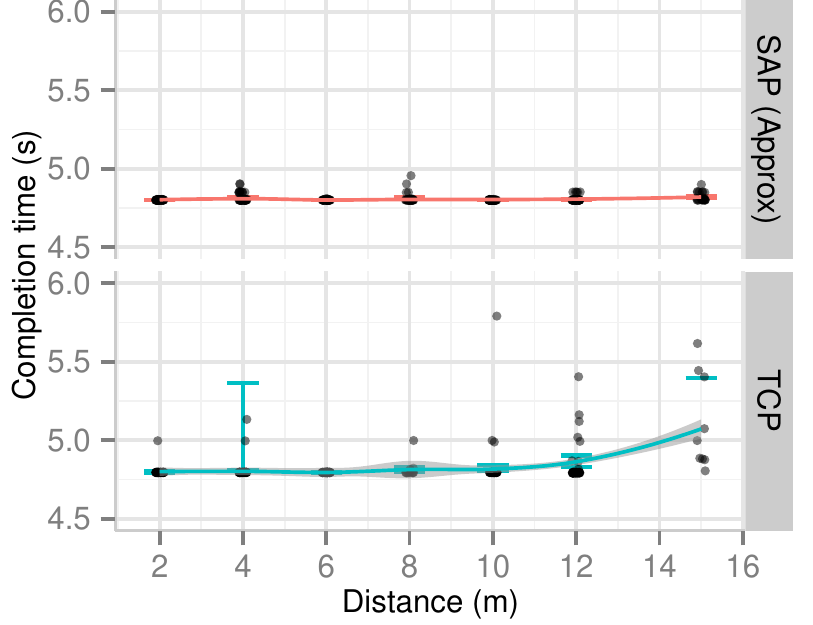}
\caption{\sysname exhibits lower variance of inter-packet arrival time than TCP.
Error bars are centered around the mean and represent standard error over $20$
trials per distance.}
\label{fig:qualtime-distance}
\end{figure}

To measure the amount of delay due to retransmissions, we reduced the delay
between the sender's transmissions to $5\,\text{ms}$; this makes long delays
figure prominently into the total time to transmit the trace.  In a tracking
application, long delays may introduce large errors when transmit queues fill
with stale data that no longer represents the true location.
As in the previous experiments, \sysname and TCP achieved similar transfer times
under good channel conditions (i.e., at close range).  As channel quality
degraded, \sysname exhibited lower variance, as in other
experiments; \figref{fig:qualtime-distance} summarizes.

As described above, the Tracker application performs online quality monitoring
with a simple sanity check on each incoming GPS reading: if the calculated speed
is beyond a reasonable bound, it discards the current distance and speed figures
and attempts to interpolate later on.  This interpolation tends to overestimate
the speed but serves its purpose as a proof of concept.
Writing application-specific quality metrics can be subtle~\cite{enerj-pldi11},
but it is a useful exercise beyond approximate computing: any operation that
degrades output quality, such as compression, can also benefit.

\section{Related Work}\label{sec:related}

\sysname is related to other systems that have proposed to allow errors to
propagate to applications.  In addition to the related work on approximation
mentioned in \secref{sec:motivation}, several schemes from the networking
literature are particularly relevant.  We focus especially on schemes that
question the importance of perfectly precise delivery.

\mypara{Multimedia streaming.}
UDP-Lite~\cite{rfc3828, Larzon99udplite}, a variant of UDP with adjustable
checksum coverage, \emph{nearly} suffices to provide end-to-end transmission of
damaged information.  Its designers point out that link-layer checksums---such
as the 802.11 FCS---counteract UDP-Lite's benefits by effectively refusing to
carry damaged data; they suggest disabling link-layer checksum checking in
device drivers to allow damaged data to pass.
\sysname builds on this suggestion by coordinating the activities of the link
layer and transport layer \emph{and} providing precise communication with
complete checksum coverage and guaranteed in-order delivery.

Singh \etal~\cite{udplite-cellular} evaluate UDP-Lite for video transmission on
cellular links that use the \emph{PPP} protocol at the link layer.  They
implement a ``PPP Lite'' mechanism that allows receivers to ignore link-layer
checksums and pass damaged data to the application layer.  Their
technique improved user-perceptible application metrics while streaming video
using error-resilient codecs---most notably the inter-arrival time of video
frames.  Servetti \etal~\cite{servetti} and
Meriaux and Kieffer~\cite{meriaux:udplite} make similar observations for speech
streaming.
\iffalse
%
These results highlight an interesting property of some error-tolerant
applications: sometimes receiving potentially damaged data \emph{now} is
preferable to receiving perfect data \emph{later} because the latter option
would create larger discontinuities.
Masala \etal~\cite{masala2004link} extend
the idea of Singh \etal by adapting link-layer retransmission behavior: only
frames with damage in \emph{consequential} payload data are retransmitted.
\fi
%
\sysname would be equally useful for the video and audio streaming tasks and
and also generalizes to other
network applications.

Khayam \etal~\cite{Khayam2003575} evaluate (in simulation) a \sysname-like
strategy for multimedia streaming over 802.11b links, with a goal of
understanding the distribution of errors that arise as a result of using
UDP-Lite.  As in \sysname, they consider removing integrity checks at low layers
to pass potentially damaged data to applications, and they observe throughput
improvements with UDP-Lite versus UDP.  They advocate using FEC at the
application layer, i.e., by encoding content at one end and decoding it at the
other, and suggest that UDP-Lite requires less FEC overhead than plain
integrity-protected UDP.  This technique complements the one we propose
for \sysname applications; application-layer logic invariably works at a
higher semantic level.

\iffalse % omit for space
Jakubczak and Katabi design Softcast~\cite{softcast}, a ``one size fits all''
transmission mechanism for grayscale video that uses specific properties of a
video-encoding scheme to allow receivers with variable channel quality to
receive the same stream with gracefully degrading fidelity.
\fi

\mypara{Recovering from damage.}
Some systems have focused on reducing retransmissions by limiting
the scope of errors and requesting only small portions of erroneous packets.
All of these techniques assume that every bit in every transmission is precious.
This assumption underlies the design of even the lowest-level primitives: even
though transport-layer protocols like UDP and UDP-Lite~\cite{Larzon99udplite}
can have their checksums disabled or limited, lower-layer 802.11 frames include
a mandatory checksum (the \term{frame check sequence}) that completely cancels
any gains from relaxing transport-layer integrity checks.

For some data formats, such as broadcast video, one way to provide
best-effort content without retransmission is to transmit the content such
that diminished channel quality results in only slight degradations---rather
than the choppy playback that conventionally accompanies packet loss.
Softcast~\cite{softcast} embodies this approach for video.  Redundant coding
schemes for other kinds of data (e.g., JPEG2000 images~\cite{frescura:jpeg2000}
or MJPEG2000 video~\cite{guo:fec}) to increase error tolerance are an open
research area.
\sysname is independent of such schemes, relying instead on error tolerance at
higher layers; applications may apply their own coding schemes that
fit their purposes exactly.

Miu \etal propose \emph{multi-radio diversity} (MRD)~\cite{mrd}, which allows a
receiver with multiple radios to receive several versions of a damaged frame and
uses a selective retransmission scheme instead of 802.11's.  MRD's extra radio
coordination requires an 802.11 network to be redesigned.  \sysname is designed
for a single-radio scenario and does not have access to information that would
allow it to recover damaged frames, but it is also backward compatible with
conventional Wi-Fi networks.

Jamieson and Balakrishnan propose \emph{partial packet recovery}
(PPR)~\cite{jamieson:ppr}, a method of minimizing retransmissions that allows
receivers to use extra confidence information (``hints'') provided by the
physical layer.
With a link-layer retransmission protocol designed to use PPR's extra PHY
information, throughput under noisy channel conditions
increased by a factor of four.
Maranello extends the PPR concept from software radio to commercial 802.11
hardware and can boost UDP throughput by 30\%~\cite{maranello}.
\sysname differs fundamentally in its approach to retransmission; applications
need not request retransmission \emph{at all} if the information they receive is
``good enough'' according to their metrics.
To use PPR or Maranello in a manner complementary to \sysname, a receiver could
lower the confidence threshold it uses to decide whether to request
retransmission.

\iffalse
Google's QUIC protocol~\cite{quic} is designed for latency-sensitive
applications to use instead of TCP.  QUIC provides several services atop UDP and
attempts to minimize the impact of packet loss by constraining cryptographic
blocks and error correction to packet boundaries, meaning that packets are
independent with respect to errors.  QUIC handles packet loss but not
corruption, in part because it leans heavily on cryptographic mechanisms to
protect payloads.  \sysname works on unencrypted payloads at present, but as a
substrate for QUIC, it might provide further speedups over plain UDP when QUIC's
damage-control mechanisms can salvage most of a damaged datagram.
\fi

\iffalse % hard to fit this in now
\xxx[br]{move this:}
Question: will middleboxes that violate the end-to-end principle for speed's
sake discard \sysname traffic~\cite{extend-tcp}?
\fi

Balan \etal propose TCP~HACK~\cite{tcp-hack}, a TCP extension that implements
TCP header checksums to allow receivers to distinguish between congestion and
packet corruption.  HACK extends the \emph{partial checksum} idea of UDP-Lite to
TCP and requires relaxation of link-layer integrity checks to be useful, so it
may be a useful building block toward supporting TCP in \sysname.
Myriad other proposals attempt to improve TCP throughput under poor channel
conditions; we draw attention mainly to HACK because it is the closest in spirit
to \sysname.

\section{Discussion}\label{sec:discussion}

\iffalse
Where to apply error checking is a crucial design principle for networks.  In
wired networks, transmission errors are rare; they generally indicate a hardware
problem if they occur with significant frequency.  In wireless networks, errors
are routine.  Wireless links in use today span a wide spectrum of error
tolerance, from extremely lossy links that use heavyweight (redundant) coding
schemes---e.g., interplanetary links where retransmission is onerous---to
lightly coded links that offer higher throughput but retransmit frequently.

\sysname focuses on improving performance over 802.11 links, which fall into the
latter category because of industry attempts to achieve parity with wired
Ethernet.  In these links, retransmission due to errors is a common occurrence,
so reducing retransmission is likely to have a significant impact.  For more
heavily coded links, relaxing coding schemes is likely to be
unacceptable for a variety of reasons.
\fi

802.11 and other wireless protocols generally use forward error correction (FEC)
in the form of coding schemes with varying efficiency.  We are unaware of any
WiFi chipsets or drivers that expose such fine-grained control over FEC.  If
such support existed, \sysname could use riskier (more efficient) coding to
increase throughput for approximate data.

\mypara{Non--802.11 networks.}
A class of devices that may benefit from approximate networking is
embedded systems, in which the contrast in power consumption between radio and
CPU is stark.  For example, a $2.4\,\text{GHz}$ radio system-on-chip on a sensor
mote that is transmitting or actively listening can use nearly $40\times$ more
energy than the microcontroller that controls it~\cite{Fonseca08:OSDI}.  On
these systems, reducing the time for which radios are awake is an important
design goal.

\mypara{Recent 802.11 variants.}
Recent variants of 802.11, particularly the \textsf{ac} variant~\cite{802.11ac},
aggregate many chunks of data per transmission and allow receivers to
acknowledge only the chunks they received correctly, an approach reminiscent of
partial packet recovery (PPR)~\cite{jamieson:ppr} and
Maranello~\cite{maranello}.
Though the \sysname prototype is implemented atop 802.11b/g for simplicity,
there is no fundamental reason it is incompatible with newer variants.  The
partial-retransmission behavior of 802.11ac is still predicated on the
assumption that the receiver wants to receive packets exactly as they were sent;
relaxing this requirement may result in similar gains.

\iffalse
% Adrian points out that this \subsection is essentially proposing a library to
% link into apps -- yawn
\subsection{Extension: Automatic Approximation}\label{sec:discuss:auto}
\sysname complements existing approximation techniques that treat data as
approximate.  If a data structure (e.g., a file) carries an ``approximate'' tag
indicating that it may contain errors, an application using \sysname can observe
this tag and switch to approximate mode when transmitting the data structure.
%
Additionally, approximation techniques that transform programs at compile time
can substitute approximate semantics for precise semantics as soon as their
analysis determines that approximation is appropriate.
\fi

\iffalse
% dropped for reasons explicated below
\subsection{Extension: TCP}\label{sec:discuss:tcp}
\xxx[br]{I think this section should be dropped; we show earlier that \sysname's
precise mode is ``close enough'' to TCP's guaranteed in-order delivery that we
don't really need TCP.  But if we're going to ditch the idea of TCP, we need to
make sure its performance is close enough to TCP's.}
\begin{itemize}
 \item Risk: larger headers mean more bits to screw up
 \item Evidence points to greater probability of packet droppage
 \item ACK behavior: just ack packets as they come in, even if they're damaged
 \item OoO delivery: assume all approx packets are in order
\end{itemize}
\fi

\mypara{Limitations.}
A limitation of \sysname's approach is that some types of data are fundamentally
not error tolerant.
For example, \sysname cannot support WPA-style encrypted frames, because errors
in these frames result in completely incorrect decryption that is unrelated to
the correct plaintext.  An appealing idea is to redundantly encode ciphertext so
that symbols can be recovered despite bit flips, but we assume that decoding
these schemes will be prohibitively energy intensive for receivers.

An alternative strategy, in the spirit of QUIC~\cite{quic}, is to
support encrypted content is to break it into multiple chunks---e.g., 128-bit
AES blocks---and allow for retransmissions only of those chunks that did not
decrypt correctly.  We leave the design of an appropriate scheme to future work
so that it may be explored more fully.

Similarly, some data types are structured such that disturbing any bit will have
arbitrarily damaging effects on the output.  Highly compressed payloads are an
example of such a data type.  Other embodiments of approximate computing suffer
from similar limitations with respect to encrypted or highly compressed
data~\cite{approxstorage-micro13}.

\iffalse % paperware mentioned elsewhere (see related work e.g. Frescura, Guo)
\subsection{Future Work}\label{sec:discuss:apps}
\xxx[br]{Drop?}

Special handling for certain content types:
\begin{itemize}
 \item Don't introduce errors in first 1k (?) of image
 \item (Test this separately)
 \item SAP: send the first 1k precisely, then the rest approx
\end{itemize}
\fi

\section{Conclusion}\label{sec:conclusion}

Approximate computing is a promising technique for better performance and
energy efficiency in computer systems. However, to this point it has focused on
computation and storage. This paper presented \sysname, a cross-layer approach
to wireless networking that gives applications a principled way to trade data integrity
for better throughput and latency. An application written
to use \sysname can interleave approximate and precise network transmission
modes, and a receiver can elect to accept potentially damaged data and perform
its own optional correctness checks.

By showing that optional \emph{approximate} network semantics on a WiFi
testbed can improve several important application-layer metrics, we hope that
\sysname will facilitate exploration of similar relaxations on present and
future networks. This effort, combined with prior efforts on accuracy
trade-offs in computation and storage, leads to end-to-end approximate computing
opportunities.

{\footnotesize \bibliographystyle{acm}
\bibliography{sap}}

\end{document}